# Rapid variations of dust colour in comet 41P/Tuttle-Giacobini-Kresák

Igor Luk'yanyk[1,*], Evgenij Zubko[2], Marek Husárik[3], Oleksandra Ivanova[3,4,1],

Ján Svoreň[3], Anton Kochergin[2,5], Alexandr Baransky[1], and Gorden Videen[6,7]

[1] Astronomical Observatory, Taras Shevchenko National University of Kyiv, 3 Observatorna St., 04053, Kyiv, Ukraine

[2] School of Natural Sciences, Far Eastern Federal University, 8 Sukhanova St., Vladivostok 690950, Russia

[3] Astronomical Institute of the Slovak Academy of Sciences, SK-05960 Tatranská Lomnica, Slovak Republic

[4] Main Astronomical Observatory of National Academy of Sciences, 27 Akademika Zabolotnoho St. 03143, Kyiv, Ukraine

[5] Ussuriysk Astrophysical Observatory of Far Eastern Branch of Russian Academy of Science, Gornotaezhnoe 692533, Russia

[6] Space Science Institute, 4750 Walnut St, Suite 205, Boulder, CO 80301, USA

[7] Kyung Hee University, 1732, Deogyeong-daero, Giheung-gu, Yongin-si, Gyeonggi-do 17104, South Korea

[*] Corresponding Author

E-mail address: iluk@observ.univ.kiev.ua

Phone: +380444862691

Fax:     +380444862691

**Abstract**

We monitor the inner coma of comet 41P/Tuttle-Giacobini-Kresák searching for variations of its colour. Fast changes in colour of the comet 41P/Tuttle-Giacobini-Kresák provide important clues for better understanding of the microphysical properties of its dust. Using the 61-cm and 70-cm telescopes we measured the apparent magnitude of the comet with the *V* and *R* Johnson-Cousins filters from January 29 until April 25 of 2017. The inner coma (~2000 km) reveals fast and significant variations of colour. The most significant change was found between March 3 and 4 of 2017, when it changed from blue with a colour slope $S \approx (-10.15 \pm 3.43)\%$ per 0.1 μm to red with $S \approx (16.48 \pm 4.27)\%$ per 0.1 μm. This finding appears in good accordance with what was previously reported by Ivanova et al. (2017) for long-period comet C/2013 UQ4 (Catalina), suggesting that fast and significant variations of colour of dust could be a common feature of short- and long-period comets. We model observations of comet 41P/Tuttle-Giacobini-Kresák using the agglomerated debris particles and conclude that its inner coma consists of a mixture of at least two types of particles made of Mg-rich silicates and organics or Mg-Fe silicates.

**Highlights**

We present the results of photometric observations of comet 41P/Tuttle-Giacobini-Kresák performed with the *V* and *R Johnson-Cousins* filters from January 29 until April 25 of 2017.

The inner coma (~2000 km) reveals fast and significant variations of colour.

The most significant change was registered between March 3 and 4 of 2017, when it changed from blue with the colour slope $S \approx (-10.15 \pm 3.43)\%$ per 0.1 μm to red with $S \approx (16.48 \pm 4.27)\%$ per 0.1 μm.

Our modeling suggests that comet 41P/Tuttle-Giacobini-Kresák consists of a mixture of at least two types of particles made of Mg-rich silicates and organics or Mg-Fe silicates

## 1. Introduction

Microphysical properties of dust particles in the vast majority of comets can be inferred only using remote-sensing techniques based on the interaction of dust with the solar radiation. Such interaction manifests itself in three groups of phenomena: elastic scattering of the sunlight in visible and near-IR bands, thermal emission of the absorbed solar radiation in mid-IR, and radiation pressure that affects the motion of dust particles and, thus, their spatial distribution in the coma and tail.

Photometric colour is defined as the ratio of the scattered-radiation fluxes measured at two different wavelengths. Its strength lies in the relative nature as it depends directly upon the microphysical properties of dust particles, and it is independent of their number within the field of view. This is important because it is known that the number of dust particles forming a cometary coma is a subject of rapid and significant variations. Photometric colour is quantified with either the *colour index* or *colour slope* (e.g., A'Hearn et al. 1984); the latter sometimes is referred to as the *reflectivity gradient* (Lamy et al. 2011). The colour slope is defined as follows:

$$S = \frac{10^{0.4\Delta m} - 1}{10^{0.4\Delta m} + 1} \times \frac{20}{\lambda_2 - \lambda_1} \quad [\% \text{ per } 0.1 \text{ μm}], \tag{1}$$

where $\Delta m$ denotes the true colour index of the comet, i.e., reduced for the colour index of the Sun, and $\lambda_2$ and $\lambda_1$ are effective wavelengths of the utilized filters measured in μm ($\lambda_2 > \lambda_1$). As one can see in Eq. (1), $S$ is normalized to the wavelength difference, making it, to some extent, independent of a specific choice of the filters.

Colour slope $S$ also could be computed directly from the reflectivity of dust particles at different wavelength $\lambda$. In practice this could be done using the $Af\rho$ parameter that was investigated in numerous comets (e.g, A'Hearn et al. 1995; Moulane et al. 2018):

$$S = \frac{Af\rho_2 - Af\rho_1}{Af\rho_2 + Af\rho_1} \times \frac{20}{\lambda_2 - \lambda_1} \quad [\% \text{ per } 0.1 \text{ μm}], \tag{2}$$

$Af\rho$ is the product of the geometric albedo $A$ of cometary dust particles to the radius of the circular field of view $\rho$, and to the *filling factor f*, which is defined as the ratio of the cumulative geometric cross section of cometary dust particles to the total projected geometric cross section of the field of view (A'Hearn et al. 1995). We note that in the literature, the *Af*ρ parameter is measured in cm. Obviously, by definition, *f* and ρ take on the same values at different wavelength λ; whereas, the geometric albedo *A* is a function of λ. In addition, the geometric albedo *A* is defined at the exact backscattering with α = 0°; whereas, the vast majority of comets are observed at α ≠ 0°. The necessity to adjust the measured *Af*ρ to the exact backscattering may cause some uncertainty in the resulting values.

Photometric colour is a key light-scattering characteristic of cometary dust that has been studied in numerous comets (e.g., Jewitt & Meech 1986; Lamy et al. 2011; Ivanova et al. 2015; Zubko et al. 2015; Picazzio et al. 2019). While the colour detected in a given comet was long considered to be a persistent feature of its dust (e.g., Jewitt & Meech 1986), it was later demonstrated that the colour can change over a considerable time period (e.g., Weiler et al. 2003; Li et al. 2014) and it is a function of the phase angle α (Zubko et al. 2014). Furthermore, Ivanova et al. (2017) have recently reported on outstanding variations of the photometric colour in comet C/2013 UQ4 (Catalina), whose dust turned from red to blue colour over only a two-day period. Note, rapid variations of colour were previously found, for instance, in comet 17P/Holmes (Zubko et al. 2011), which was clearly related to its dramatic outburst activity at the end of October, 2007; whereas, comet C/2013 UQ4 (Catalina) did not reveal such activity during its observations. The findings by Ivanova et al. (2017) raise a legitimate question as to whether comet C/2013 UQ4 (Catalina) was an exceptional case in terms of fast and dramatic temporal variations of colour in its dust. In this paper we present results of our recent photometric survey of comet 41P/Tuttle-Giacobini-Kresák (hereafter 41P/T-G-K) over a three-month period in the first half of 2017. As it is shown in the next sections, photometric colour of dust in this short-period comet underwent dramatic short-term variations, very much resembling what was found in the long-period comet C/2013 UQ4 (Catalina) by Ivanova et al. (2017).

**2. Observations and reductions**

We observed comet 41P/T-G-K on 17 nights from January 29 to April 25, 2017 using the 61-cm telescope at the Skalnaté Pleso (Astronomical Institute of Slovak Academy of Science, Slovakia, Observatory Code – 056) and the 70-cm AZT-8 (observation station Lisnyky of the Astronomical Observatory of Taras Shevchenko National University of Kyiv, Ukraine, Observatory Code – 585). A detailed log of observations is presented in Table 1. We started observing 41P/T-G-K prior to its perihelion passage (April 13, 2017), at the geocentric and heliocentric distances of $\Delta \approx 0.436$ au and $r_h \approx 1.414$ au respectively, and phase angle $\alpha \approx 8.7°$.

The observations were finished at $\Delta \approx 0.177$ au, $r_h \approx 1.058$ au, and $\alpha \approx 68.1°$. To prevent bright stars from falling into the aperture we investigate only the innermost part of the coma with projected diameter of approximately 2,000 km that we kept constant throughout all observations. However, this aperture size also nearly coincides with what was used previously in Ivanova et al. (2017). Furthermore, as shown below with modeling of dust-particle motion, the population of the dust within a 2,000-km aperture can be fully replaced during only a single day. Therefore, it is the best region for searching for fast changes of microphysical properties in the expelled dust. Data overnight observations were averaged but values that deviated from the average by more than 3-$\sigma$ were discarded. Photometry was conducted with the *V* and *R* broadband filters from the *Johnson-Cousins* photometric system that are centered at $\lambda = 0.551$ μm (FWHM = 0.088 μm) and 0.658 μm (FWHM = 0.138 μm), respectively. The colour index of the Sun with this pair of filters is equal to 0.368 (Burlov-Vasiljev et al. 1995, 1998).

The CCD detector SBIG ST-10XME with 2 × 2 binning and resolution of 1.07 arcsec per pixel was used for observation at the 61-cm telescope at the Skalnaté Pleso. The CCD PL47-10FLI camera was used as a detector with full field of view of the CCD being 16 × 16 arcminutes, and the image scale was 0.95 arcsec per pixel for observations from the 70-cm AZT-8. Fig. 1 shows image of 41P/T-G-K obtained with the 70-cm AZT-8 telescope on observation station Lisnyky of the Astronomical Observatory of Taras Shevchenko National University of Kyiv (Ukraine) of 25 April 2017. The black circle marks part of the 41P/T-G-K coma whose colour was investigated in this study.

We followed the reduction procedure (bias subtraction, dark and flat field corrections, and cleaning cosmic-ray tracks) in the standard manner, using the IDL routines (e.g., Ivanova et al. 2016, 2017; Picazzio et al. 2019). The morning sky was exposed to provide a flat-field correction for the non-uniform sensitivity of the CCD chip. All of the nights were photometric, the seeing value measured as the average FWHM of several sample stars ranged from 3.1 to 6.6 arcsec during our observations. Observations with a large seeing were not taken into account. The residual sky background was estimated with the use of an annular aperture. To perform an absolute flux calibration of the comet images, the field stars were used. The stellar magnitudes of the standard stars were taken from the catalogue UCAC4. UCAC4 is a compiled, all-sky star catalogue covering mainly the 8 to 16 magnitude range in a single bandpass between *V* and *R*. Photometric uncertainty of the catalogue depends on star brightness and is estimated to be from 0.01 to 0.2 mag on average.

## 3. Results and discussion

Fig. 2 shows results of the three-month monitoring of the colour slope *S* of comet 41P/T-G-K (points). Measured values are given in Table 2. What immediately emerges from this figure is a significant dispersion of all 17 measurements. In particular, on 9 epochs, the comet was detected with confidence to be red, on 1 epoch its colour was unambiguously blue; whereas, on 7 other epochs the colour appeared to be neutral within the error bars. Such dispersion of the colour slope in comet 41P/T-G-K supports the conclusion previously drawn by Ivanova et al. (2017) that a sole observation hardly characterizes the colour of a comet.

In addition to our measurements in Fig. 2, we also plot the colour slope *S* computed using Eq. (2) from the *Afρ* parameter measured by Moulane el al. (2018) using a 10,000-km aperture (ρ = 5,000 km) in comet 41P/T-G-K (open circles). In particular, we adapt their data obtained with the HB narrow-band blue continuum (*BC*, λ = 0.445 μm) and red continuum filters (*RC*, λ = 0.715 μm) as they cover a relatively long time period, 14 epochs, which is consistent with our own observations. Note, they also exploited the HB green continuum filter (*GC*, λ = 0.525 μm) that better corresponds to the *V* filter in our observations. Unfortunately, the *GC* filter was used only a few times. Nevertheless, we involve those data into a later analysis.

Although Moulane et al. (2018) draw attention to the colour variations in 41P/T-G-K, their analysis is largely qualitative as they do not provide the colour slope *S*. As one can see in Fig. 2, both sets of data were obtained at similar uncertainty, and within the error bars, they reveal a great deal of overlap. However, some of our data, such as the strong blue colour on March 3, 2017, appear to be unique.

In Fig. 3 we address the colour slope *S* obtained solely in our own observations. Here it is shown as a function of phase angle α. As one can see in Fig. 3, the average colour slope $S \approx$ 3.81% per 0.1 μm shown with the horizontal short-dashed line is not representative for most of observations. Only 7 out of 17 observations appear in accordance with the average colour slope within the error bars. The dramatic change of the colour slope that occurred between the 3$^{rd}$ and 4$^{th}$ of March (phase angle about 25°) is also worth mentioning. While on March 3, comet 41P/T-G-K revealed a strong blue colour with $S \approx (-10.15 \pm 3.43)$% per 0.1 μm, only one day later, it was found to be very red in appearance, with $S \approx (16.48 \pm 4.27)$% per 0.1 μm. This change is even stronger and faster compared to what was previously reported for comet C/2013 UQ4 (Catalina) by Ivanova et al. (2017). However, our current target, comet 41P/T-G-K, belongs to the class of short-period comets; whereas, comet Catalina is a long-period comet. This suggests that the fast and significant colour variations could be a common feature of comets regardless of their orbits.

An important question with regard to the colour variations in the inner coma of comet 41P/T-G-K is whether such variations should be attributed to changes in the dust population or gaseous environment. In theory, gaseous emission potentially could affect photometric measurements conducted with the wideband filters. In practice, however, such a gaseous contamination does not appear to be a major factor. This was checked, for instance, by integrating high-resolution cometary spectra in visible with and without gaseous-emission lines taken into account. The difference between those two types of spectral integrations exceeds only a few percent (e.g., Jewitt et al. 1982; Picazzio et al. 2019), suggesting an extremely low contribution of the gaseous emission into the flux measured with wideband filters. Interestingly, this equally holds for long- and short-period comets. One also should draw attention to comparative analysis of the color slope in Comet C/1995 O1 (Hale-Bopp) inferred from its high-resolution spectra and, simultaneously, from photometry with the wideband *B* and *R* filters. As reported in Weiler et al. (2003), both approaches yield nearly the same result within the error bars, despite an outstanding gaseous emission in this comet. Unfortunately, high-resolution spectra of comets in the visible are not always available for validation of the true gaseous-emission contribution into their photometric response measured with the wideband filters. It often gives rise to speculation.

Nevertheless, the gaseous-emission contribution into the wideband-filter photometric response of comet 41P/T-G-K could be estimated from a comparative analysis of production rates of its gases Q and the *Af*ρ parameter of its dust. While Q refers to the number of molecules of a certain gas produced from the nucleus per second, the *Af*ρ parameter provides a quantitative estimation of the abundance of dust particles in the coma.

The production rates of some gases and the *Af*ρ parameter of dust in comet 41P/T-G-K on its 2017 apparition were reported recently in Moulane et al. (2018). For instance, on the earliest available dates, March 11–27 of 2017, the OH production rate, the daughter species of $H_2O$, ranged from Q(OH) ≈ 219 × $10^{25}$ molecules $s^{-1}$ to 257 × $10^{25}$ molecules $s^{-1}$. On the same epochs, the production rate of the $C_2$ molecule, whose fluorescence could potentially affect the signal measured with the wideband Johnson *V* filter, was found to be Q($C_2$) ≈ (0.38 – 0.62) × $10^{25}$ molecules $s^{-1}$. Simultaneously, the *Af*ρ parameter remained nearly constant in the blue continuum (BC) filter, *Af*ρ = 54.7 – 56.2 cm. These data make it possible to compute two important characteristics, log(*Af*ρ/Q(OH)) ≈ – (25.63 ± 0.03) and log(Q($C_2$)/Q(OH)) ≈ – (2.66 ± 0.1), that quantify the relative abundance of dust particles and the $C_2$ molecules with regard to the OH molecules in comet 41P/T-G-K. Clearly, these characteristics alone do not provide clues sufficient for understanding the relative strength of the gaseous emission. However, these characteristics have already been studied in numerous comets (e.g., A'Hearn et al. 1995), including comets C/1980 Y2 (Panther) and 38P/Stephan–Oterma. In the former case, A'Hearn et al. 1995) reported $\log(Af\rho/Q(OH)) = -25.32$ and $\log(Q(C_2)/Q(OH)) = -2.35$; whereas, in the latter one, $\log(Af\rho/Q(OH)) = -25.47$ and $\log(Q(C_2)/Q(OH)) = -2.80$. Thus, comet 41P/T-G-K appears in good accordance with C/1980 Y2 (Panther) and 38P/Stephan–Oterma. What is significant is that the comets C/1980 Y2 (Panther) and 38P/Stephan–Oterma were also investigated with the spectroscopic approach by Jewitt et al. (1982). Spectra of both comets in the visible reveal strong emission lines from various gases. However, integration of spectra over the wavelengths $\lambda = 0.5 - 0.7$ µm with and without those emission lines reveals only little difference, suggesting an extremely weak contribution of the gas emission, namely, a gaseous emission contribute only 1% of the total flux in comet C/1980 Y2 (Panther) and even less in 38P/Stephan–Oterma (Jewitt et al. 1982). Therefore, we conclude that our photometric measurements with the *V* and *R* filters correspond primarily to dust in comet 41P/T-G-K.

As can be seen in Fig. 2, colour variations in comet 41P/T-G-K detected in our study with the broadband *V* and *R* filters appear overall consistent with what was found in this comet on the same apparition with the *HB BC* and *RC* filters by Moulane et al. (2018). However, their continuum filters *HB GC* ($\lambda = 0.525$ µm) and *RC* ($\lambda = 0.715$ µm) are somewhat better embraced by bandpasses of our *V* and *R* filters, making possible a comparison of the results obtained with both sets of filters. Using the pair of *GC* and *RC* filters, Moulane et al. (2018) measured comet 41P/T-G-K on four epochs. Two of them, 2017-03-26.04 and 2017-04-02.88, nearly coincide with our measurements; see Table 2. We compute the color slope *S* using values of the *Af*ρ parameter presented in Table 2 of Moulane et al. (2018) for the *GC* and *RC* filters, and find $S = (7.87 \pm 3.30)$% per 0.1 µm for former date and $S = (9.87 \pm 2.24)$% per 0.1 µm for the later epoch. These results match our finding within the error bars (see Table 2) despite that the measurements by Moulane et al. (2018) correspond to a bigger projected area of the coma, a radius of 5,000 km versus ~1,000 km in our measurements. Unfortunately, Moulane et al. (2018) does not report the continuum photometric data on the night between March 3 and 4, making our observations on those epochs unique.

We analyze the observations presented in Fig. 3 using the model of the *agglomerated debris particles*. These particles have highly irregular and fluffy morphology with bulk material density ranging from 0.35 g/cm$^3$ to 0.83 g/cm$^3$. These features appear in good accordance with what was found in cometary and interplanetary dust particles *in situ* (see, e.g., Section 3 of Zubko et al. 2016). Five examples of the agglomerated debris particles are shown at the top of Fig. 4; whereas, at the bottom, we reproduce images of the micron-sized particles sampled by *Rosetta* in the vicinity of 67P/Churyumov–Gerasimenko nucleus (adapted from Bentley et al. 2016). As one can see, the shape of the agglomerated debris particles resembles that of the cometary dust particles. It is significant that the agglomerated debris particles reveal an outstanding capability to reproduce photometric and polarimetric observations of various comets (see, e.g., Zubko et al. 2011, 2014, 2015, 2016; Ivanova et al. 2017; Picazzio et al. 2019) as well as laboratory optical measurements of cometary-dust analogs (Zubko 2015; Videen et al. 2018). We refer the reader to works of Zubko et al. (2015), Ivanova et al. (2017), Picazzio et al. (2019), where the modeling of the colour slope with the agglomerated debris particles is described in detail. In Fig. 5, we largely adapt the modelling results previously obtained in those works, adjusting the phase angle to α = 25° that comet 41P/ T-G-K had on March 3 – 4 of 2017, when its colour dramatically changed. Interestingly, on March 3 and 4, the colour slope $S$ has reached minimum and almost maximum values over the entire period of our observations.

We note that the colour of cometary dust also depends on phase angle α. For example, in Fig. 3 we show with purple solid and long-dash lines the colour slope as a function of phase angle inferred in comet C/1975 V1 (West) within two different scenarios (data adapted from Zubko et al. 2014). Nevertheless, day-to-day variations of $S$ apparently may significantly exceed those, at least, in the given range of phase angle.

The upper plot in Fig. 5 shows the colour slope $S$ of agglomerated debris particles as a function of the index $n$ in their power-law distribution $r^{-n}$ at nine refractive indices $m$ that are representative of various refractory species in comets (see, e.g., Zubko et al. 2015). We here assume that there is no wavelength dependence of the imaginary part of refractive index, i.e., $\Delta \mathrm{Im}(m) = 0$. Two striped purple rectangles mark the colour slopes measured in comet 41P/T-G-K on March 3 and 4 with their error bars. As one can see, none of the refractive indices $m$ is capable of producing the very red colour detected on March 4; whereas, the blue colour on March 3 can be fitted at some values of the power index $n$. Two outstanding cases here are $m$ = 1.313 + 0$i$ and 1.6 + 0.0005$i$, which correspond to water ice and Mg-rich silicates, two known species of comets. They indeed reveal no wavelength dependence of their $m$ in visible (Warren 1984; Dorschner et al. 1995). Moreover, both appear consistent with our observation on March 3 at a very wide range of power index $n$ and, simultaneously, at $n \leq 3$ that makes them also consistent with the *in situ* finding on the size distribution of submicron and micron-sized particles in comets (Mazets et al. 1986; Price et al. 2010). Although agglomerated debris particles at other refractive indices may also reproduce $S$ on March 3, they suggest $n > 3$ that is beyond the bounds suggested by these *in situ* studies in micron-sized dust particles. The same hold for the results presented in the middle plot ($\Delta \mathrm{Im}(m) = 0.01$) and on the bottom plot ($\Delta \mathrm{Im}(m) = 0.02$). Therefore, we rule out all those fits to $S$ on March 3. Furthermore, due to the small heliocentric distance of 41P/T-G-K, $r_h \approx 1.18$ au, the presence of micron-sized water-ice particles in its coma seems very unlikely. Thus, we conclude that on March 3, the coma consisted mainly of Mg-rich silicates with power index constrained to the range $n \approx 2.2 - 3$.

The colour slope measured on March 4 unambiguously suggests some wavelength dependence of the refractive index. In the middle and bottom plots we show modeling results obtained by varying the imaginary part of refractive index Im($m$). We assume that at the shorter wavelength, Im($m$) takes on a larger value than at the longer wavelength; whereas, the real part of refractive index Re($m$) remains constant. Such a trend is common for cometary-dust species in visible (see, discussion in Zubko et al. 2015). One also needs to note that, unlike Im($m$), a small variance of Re($m$) does not significantly affect the light-scattering response of irregularly shaped particles (Zubko et al. 2014).

As one can see in the middle plot of Fig. 5, at $\Delta$Im($m$) = 0.01, all the modelling scenarios are capable of fitting the observation on March 4 at $n < 2.3$. What is significant is that some of these cases (i.e., Im($m$) = 0.02 – 0.03 at shorter wavelength $\lambda(S)$ and Im($m$) = 0.01 – 0.02 at longer $\lambda(L)$) can be attributed to real materials like organics and Mg-rich silicates with some contamination from iron (Jenniskens 1993; Dorschner et al. 1995; Zubko et al. 2015).

The reddening further increases at $\Delta$Im($m$) = 0.02, see the bottom plot in Fig. 5. We discriminate here two solutions obtained with Im($m$) = 0.04 – 0.06 at shorter $\lambda$ and Im($m$) = 0.02 – 0.04 at longer $\lambda$. They not only appear consistent with some organics and Mg-Fe silicates, but also fit the colour slope $S$ on March 4 with the power index $n \approx 2 - 2.9$ that nearly coincides with what was inferred with Mg-rich silicates for March 3. Thus, within these scenarios both components of the 41P/T-G-K coma would obey the same size distribution.

As for a mechanism governing the replacement of pure Mg-rich silicates on March 3 with organics/Mg-Fe silicates on March 4 in the inner part of the 41P/T-G-K coma, we simulated the motion of dust particles within the model previously exploited in Zubko et al. (2015) and further expanded in Kochergin et al. (2018). The model is based on a time-domain approach of computation of dust-particle motion. At the initial time moment $t_0$, a dust particle gets lifted off the nucleus surface at the place of its origin characterized with the radius-vector $\mathbf{s}_0$ and acquires initial velocity $\mathbf{V}_0$. We adapt $R = 0.7$ km for the radius of the 41P/T-G-K nucleus from Lamy et al. (2004). In general, the particle motion is affected by three forces, the solar gravity $\mathbf{F}_{Sun}$, the cometary nucleus gravity $\mathbf{F}_{nucleus}$, and the solar-radiation pressure $\mathbf{F}_{rad}$; the latter one acts in the direction opposite to the solar gravity (e.g., Fulle 2004). We infer the gravitational force of the 41P/T-G-K nucleus based on the assumption of its bulk material density of 0.5 g/cm$^3$ that is within the range detected *in situ* in cometary nuclei (Weissman & Lowry 2008).

The resultant of all three forces changes the dust-particle motion with the acceleration $\mathbf{a}_0$. We hold $\mathbf{a}_0$ constant during the first second that suggests a uniformly accelerated straight-line motion of the dust particle over the corresponding time period. At the end of the first time step, we update the particle velocity $\mathbf{V}_1$ and all three forces at its next location at $\mathbf{s}_1$, which yields a new value of their resultant acceleration $\mathbf{a}_1$. We repeat this whole computation procedure with time increment $\Delta t = 1$ s for a full day, i.e., 86,400 s. What emerges from this simulation is a sequence of dust-particle locations composing its trajectory. Note that we examine different values of the time increment and find that $\Delta t = 1$ s produces a well-converged modeling result. We also consider at every particle location whether it is shadowed by the nucleus from the direct solar radiation. If the particle is shadowed, we omit the solar-radiation pressure during the next increment of time $\Delta t$.

In the literature, the effect of the solar radiation on dust-particle motion is quantified with the $\beta$ parameter that is a ratio of the solar-gravity force over the radiation-pressure force $\beta = F_{Sun} / F_{rad}$ (e.g., Fulle 2004). $F_{rad}$ is dependent on the radiation pressure efficiency $Q_{pr}$ whose computation requires a rigorous solution of the light-scattering for the given dust particle (e.g., Bohren & Huffman 1983). In application to dust particles with truly irregular morphology this can be accomplished only using a highly time-consuming numerical approach, such as, the discrete dipole approximation (DDA) (e.g., Zubko et al. 2015). This necessity is often ignored, and $Q_{pr}$ often gets preset to 1 instead. The reasoning behind this approximation is the asymptotic behavior $Q_{pr} \to 1$ as the particle radius $r >> 1$ µm (e.g., Fulle 2004; Moreno et al. 2012), except for water-ice particles (Zubko et al. 2015). This approximation was recently utilized, for instance, in application to comet 41P/T-G-K by Pozuelos et al. (2018), where the shape of its coma was modeled with particles whose radius $r$ spans the range from 10 µm to 1 cm. However, the contribution of such large particles into the light-scattering response from a cometary coma is not established. While large particles with radius in excess of 10 µm do indeed exist in a cometary coma (e.g., Bentley et al. 2016), there also are present submicron- and micron-sized dust particles in much greater quantities (e.g., Mazets et al. 1986; Price et al. 2010), and these smaller particles are more efficient scatterers. As demonstrated by Zubko (2013), if the index in the power-law size distribution $r^{-n}$ ranges from $n = 1.5 – 3.5$, the light-scattering at phase angles $\alpha \leq 30°$ is predominantly governed by particles with radius $r \leq 1.5$ µm. This apparently holds for comet 41P/T-G-K on March 3–4 of 2017, when its phase angle was $\alpha \approx 24.8–26°$.

It is worth noting that when $n > 3$, which also includes the range $n = 3.5 – 3.75$ suggested by Pozuelos et al. (2018), the total geometric cross section of the dust population is governed by the bottom limit of integration. However, in the analysis by Pozuelos et al. (2018) the bottom size of dust particles cannot be less than 10 μm in principle due to their approximation $Q_{pr}$ = 1. This contradiction makes their analysis difficult to reconcile with application to interpreting observations of 41P/T-G-K in the visible. In our modeling we exploit rigorous values of the β parameter in the Mg-rich silicate particles adapted from Zubko et al. (2015). It is significant that they are based on numerically exact DDA computations of $Q_{pr}$ in the agglomerated debris particles and integrated over the solar spectrum. In this work, we model the motion of Mg-rich silicate particles having radius $r$ = 1 μm with β = 0.5.

In Fig. 6 we show trajectories of the Mg-rich silicate particles ejected from comet 41P/T-G-K at four different velocities: 10 m/s (panel A), 20 m/s (B), 40 m/s (C), and 80 m/s (D). Here the nucleus is placed at the origin of the coordinate system, with the Sun located at the top at heliocentric distance of the comet $r_h$ ≈ 1.18 au that corresponds to March 3 of 2017. The direction to the observer (i.e., Earth) is inclined from the direction to the Sun by α ≈ 24.8° and shown with an arrow marked E. We consider an isotropic ejection of dust particles from the nucleus. The length of each trajectory corresponds to a one-day flight of the dust particle; except for the panel D, where the trajectories are clipped at the bottom side of the panel. The grey rectangle imposed over the trajectories is oriented by its short side being perpendicular to the line of sight; whereas, the length of the short side is about 2000 km and corresponds to the aperture in our measurements. The particle contributes to our measurements when its trajectory lies within this rectangular.

As one can see, the slowest-ejected particles (panel A) will totally leave the field of our view within a time period considerably shorter than one day. This occurs because the radiation pressure quickly pushes back the slowly ejected particles. The time period that dust particles spend within our field of view increases with the ejection velocity. Nevertheless, at $V_0$ = 20 m/s, all the particles ejected from the 41P/T-G-K nucleus on March 3 should still get swept out by radiation pressure by March 4. However, at $V_0$ = 40 m/s, this one-day time period is not sufficiently long to remove all the Mg-rich silicate particles by March 4, as a considerable fraction remains within the 2000-km aperture. Interestingly, further increases of the ejection velocity again decrease the fraction of the Mg-rich silicate particles, as seen for the case of $V_0$ = 80 m/s demonstrated in panel D. Thus, we conclude that the Mg-rich silicate particles observed on March 3 should have left the inner coma (projected diameter of ~2000 km) within only 1 day if their ejection velocity either is smaller than 20 m/s or exceeds 120 m/s. The latter constraint seems reasonable for submicron and micron-sized dust particles ejected from a comet whose heliocentric distance is ~1.18 au. It also is worth noting that in March of 2017, the 41P/T-G-K nucleus had a 25-*h* rotational period (Bodewits et al. 2018); whereas, the time difference between observations on March 3 and 4 was ~22.6 *h*. In other words, the time difference was shorter than the full-turn period of the nucleus. This could explain the absence of the Mg-rich silicate particles in the innermost coma on March 4. Alternatively, expelling the Mg-rich silicate particles from the nucleus could occur irregularly. Finally, one can consider the chemical compositions of the coma deduced for March 3 and 4 as two different materials in the 41P/T-G-K coma; whereas, all the intermediate values of the colour slope *S* then could be considered as a mixture of those two end-members.

**4. Conclusion remarks**

A photometric survey of comet 41P/T-G-K shows fast and dramatic variations of colour in its inner coma. During only one day, between March 3 and 4 of 2017, the colour changed from blue to red. Having found such variations in a second comet suggests that such variations of colour might not be an uncommon feature. Our modelling with the agglomerated debris particles suggests that the 41P/T-G-K coma consists of a mixture of at least two end-members, Mg-rich silicates and organics/Mg-Fe silicates. These refractory species are in accordance with *in situ* studies of comets. Furthermore, both components obey the same power-law size with the index *n* ≈ 2 – 3 that also agrees with in situ findings of submicron and micron-sized dust particles in comets.

**Acknowledgements**

IL thanks the SAIA Programme for financial support. OI thanks the SASPRO Programme No. 1287/03/01 for financial support. The research leading to these results has received funding from the People Programme (Marie Curie Actions) European Union's Seventh Q12 Framework Programme under REA grant agreement No. 609427. Research has been further co-funded by the Slovak Academy of Sciences grant VEGA 2/0023/18. Also this article was created by the realisation of the project ITMS No. 26220120029, based on the supporting operational Research and development program financed from the European Regional Development Fund. The authors thank Dr. Venkataramani for discussion on gaseous emission in comet 41P/T-G-K and Dr. Bodewits for valuable comments on this manuscript.

**References**

A'Hearn, M. F., Schleicher, D. G., Millis, R. L., Feldman, P. D., & Thompson, D. T. 1984, AJ, 89, 579

Bentley, M. S., Schmied, R., Mannel, T., et al. 2016, Nature, 537, 73

Bodewits, D., Farnham, T. L., Kelley, M. S., & Knight, M. M. 2018, Nature, 553, 186

Bohren C. F., Huffman D. R., 1983, Absorption and Scattering of Light by Small Particles. Wiley, New York, USA

Burlov-Vasiljev, K. A., Gurtovenko, E. A., & Matvejev, Y. B. 1995, Sol. Phys., 157, 51

Burlov-Vasiljev, K. A., Matvejev, Y. B., & Vasiljeva, I. E. 1998, Sol. Phys., 177, 25

Dorschner, J., Begemann, B., Henning, T., Jaeger, C., & Mutschke, H. 1995, A&A, 300, 503

Fulle M. 2004, in Festou M. C., Keller H. U., Weaver H. A., eds, Comets II. Univ. Arizona Press, Tucson, p. 565

Ivanova, O., Neslušan, L., Krišandová, Z. S., et al. 2015, Icarus, 258, 28

Ivanova, O. V., Luk'yanyk, I. V., Kiselev, N. N., et al. 2016, Planet. Space Sci., 121, 10

Ivanova, O., Zubko, E., Videen, G., et al. 2017, MNRAS, 469, 2695

Jenniskens, P. 1993, A&A, 274, 653

Jewitt, D. C., Soifer, B. T., Neugebauer, G., Matthews, K., & Danielson, G. E. 1982, AJ, 87, 1854

Jewitt, D., & Meech, K. J. 1986, ApJ, 310, 937

Kochergin A., Zubko E., Popel S.I., & Videen, G. 2018, In The Ninth Moscow Solar System Symposium, (October 8–12, 2018; Moscow, Russia), contribution 9MS3-SB-04, 106

Lamy P. L., Toth I., Fernandez Y. R., & Weaver H. A. 2004, in Festou M. C., Keller H. U., Weaver H. A., eds, Comets II. Univ. Arizona Press, Tucson, p. 223

Lamy, P. L., Toth, I., Weaver, H. A., A'Hearn, M. F., & Jorda, L. 2011, MNRAS, 412, 1573

Li, J.-Y., Samarasinha, N. H., Kelley, M. S. P., et al. 2014, ApJ, 797, L8

Mazets, E. P., Aptekar, R. L., Golenetskii, S. V., et al. 1986, Nature, 321, 276

Moreno F., Pozuelos F., Aceituno F., et al. 2012, ApJ, 752, 136

Moulane, Y., Jehin, E., Opitom, C., et al. 2018, A&A, 619, A156

Picazzio, E., Luk'yanyk, I. V., Ivanova, O. V., et al. 2019, Icarus, 319, 58

Pozuelos F. J., Jehin J, Moulane Y., et al. 2018, A&A, 615, A154

Price, M. C., Kearsley, A. T., Burchell, M. J., et al. 2010, Meteoritics and Planetary Science, 45, 1409

Videen, G., Zubko, E., Arnold, J. A., et al. 2018, J. Quant. Spectr. Rad. Transf., 211, 123

Warren, S. G. 1984, Appl. Opt., 23, 1206

Weiler, M., Rauer, H., Knollenberg, J., Jorda, L., & Helbert, J. 2003, A&A, 403, 313

Weissman P. R., & Lowry, S. C. 2008, Meteoritics Planetary Sci., 43, 1033

Zubko, E. 2013, Earth Planets Space, 65, 139

Zubko, E. 2015, Optics Letters, 40, 1204

Zubko, E., Furusho, R., Kawabata, K., et al. 2011, J. Quant. Spectr. Rad. Transf., 112, 1848

Zubko, E., Muinonen, K., Videen, G., & Kiselev, N. N. 2014, MNRAS, 440, 2928

Zubko, E., Videen, G., Hines, D. C., et al. 2015, Planet. Space Sci., 118, 138

Zubko, E., Videen, G., Hines, D. C., & Shkuratov, Y. 2016, Planet. Space Sci., 123, 63

**Figures**

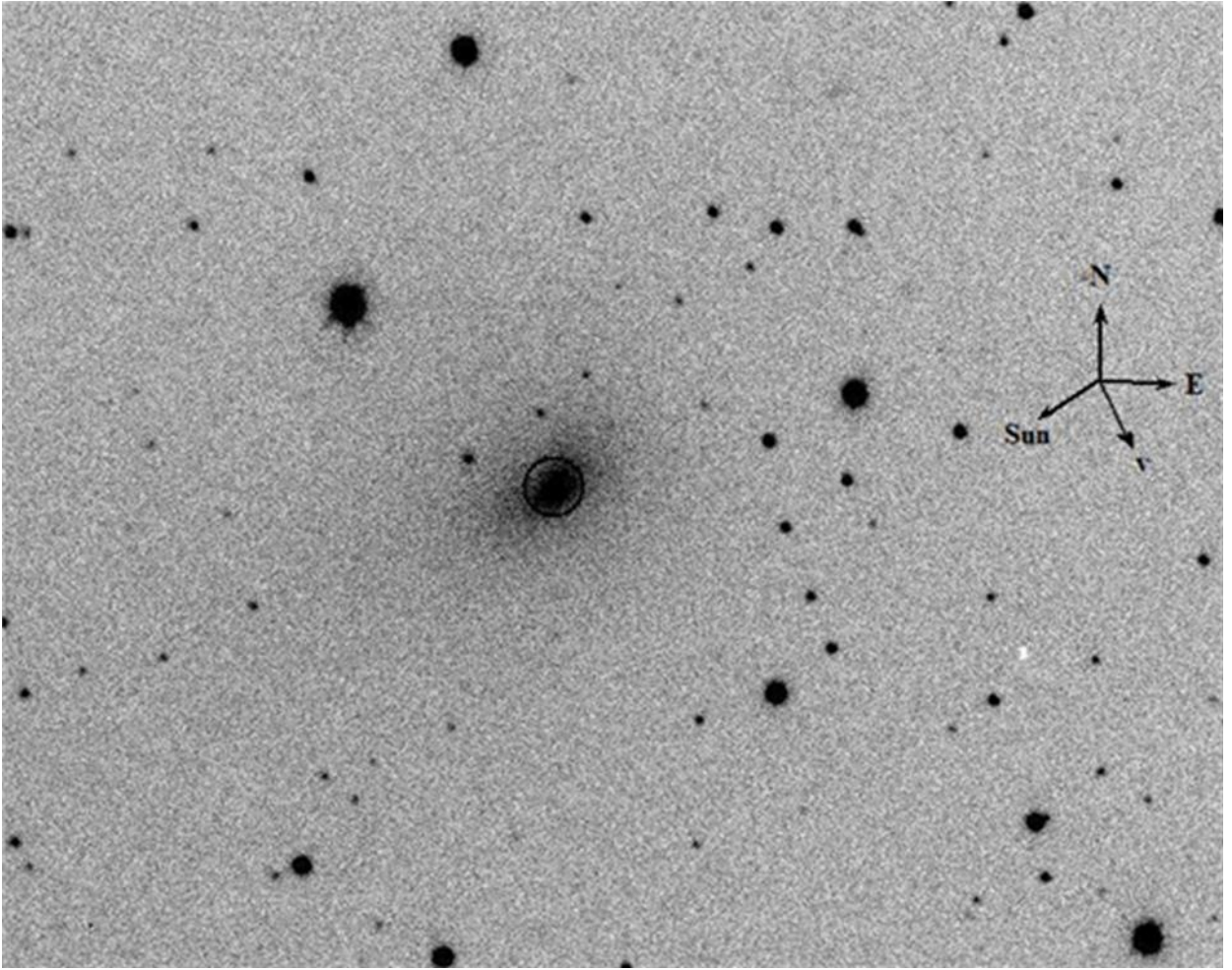

**Figure 1:** Image of 41P/T-G-K obtained with the 70-cm AZT-8 telescope on observation station Lisnyky of the Astronomical Observatory of Taras Shevchenko National University of Kyiv (Ukraine) of 25 April 2017 at the geocentric and heliocentric distances of $\Delta \approx 0.18$ AU and $r_h \approx 1.06$ AU respectively, and phase angle $\alpha \approx 68.1°$. The picture was taken with 30 second exposure through a red filter from the *Johnson-Cousins* photometric system. Also shown are schematics for the observing geometry on this date and aperture size (about 2000 km) for photometry measurements.

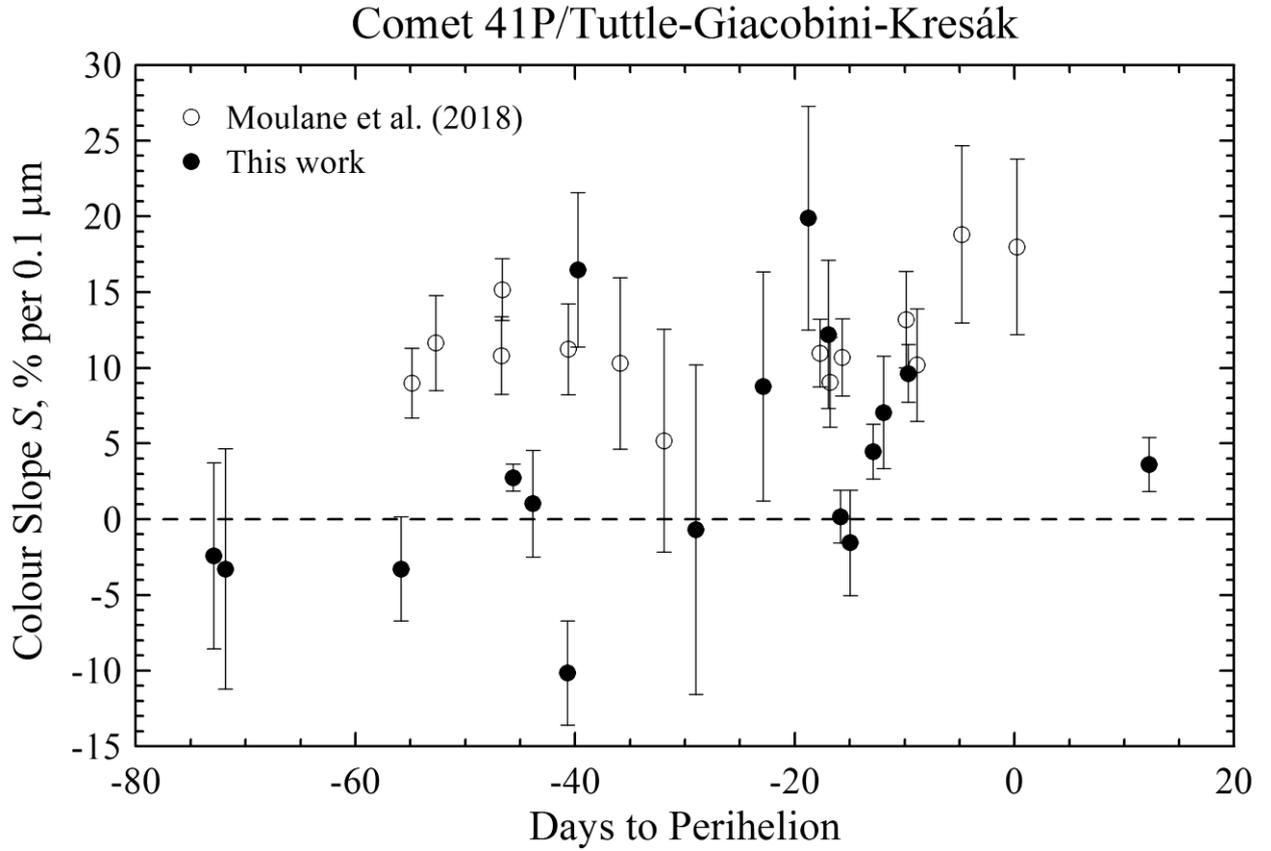

**Figure 2:** Colour slope *S* of comet 41P/T-G-K on different epochs in 2018. Results adapted from Moulane et al. (2018) were obtained with the HB narrow-band blue continuum (BC) and red continuum (RC) filters; whereas, our measurements were conducted with the Johnson-Cousins wideband *V* and *R* filter. Size of aperture in Moulane et al. (2018) is five times larger, 10,000 km versus 2,000 km in our observations.

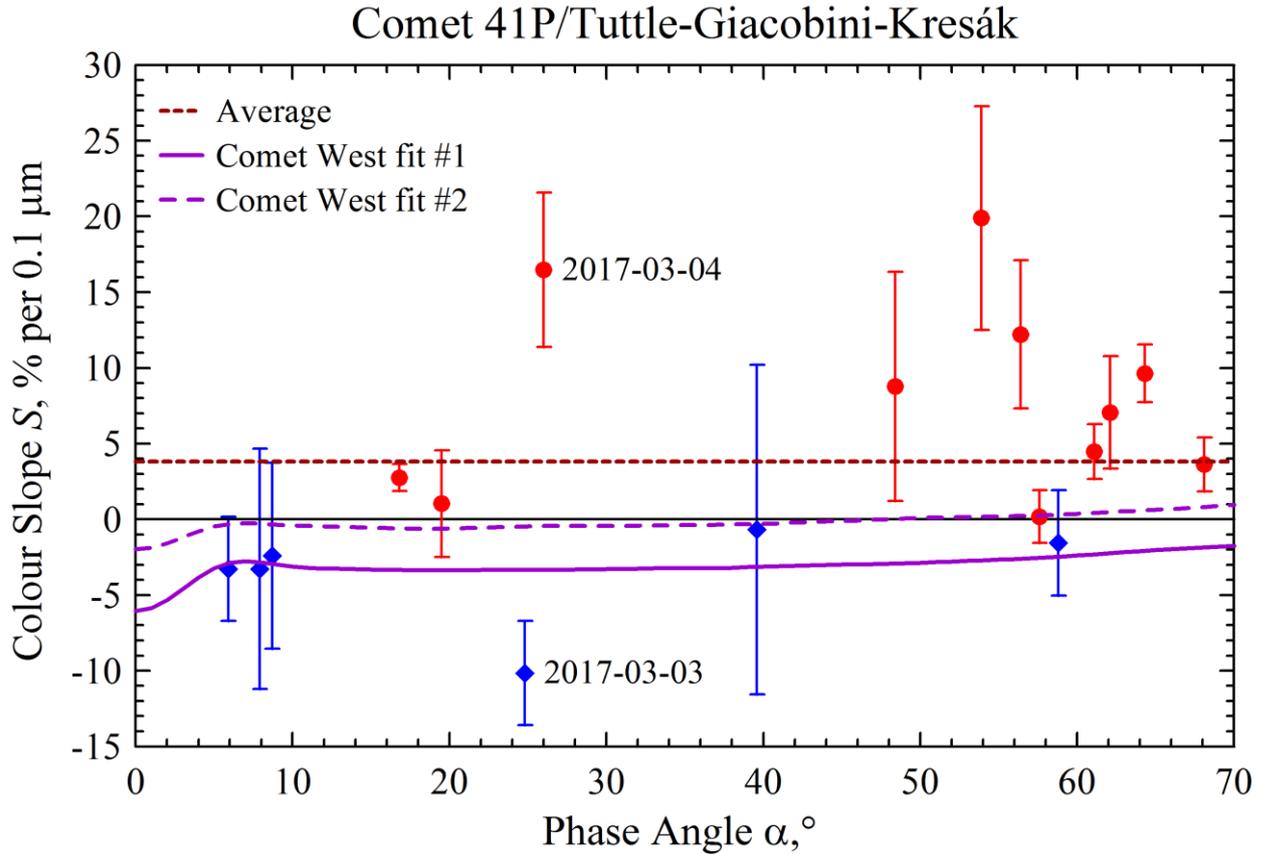

**Figure 3:** Colour slope $S$ of comet 41P/T-G-K as a function of phase angle α. Red points and blue diamonds correspond to positive and negative values of $S$, respectively. The brown short-dash line shows the colour slope average over all 17 observations. The purple solid and long-dash lines show the colour of dust in comet C/1975 V1 (West) inferred from simultaneous fitting of its phase function and angular profile of the degree of linear polarization (Zubko et al. 2014).

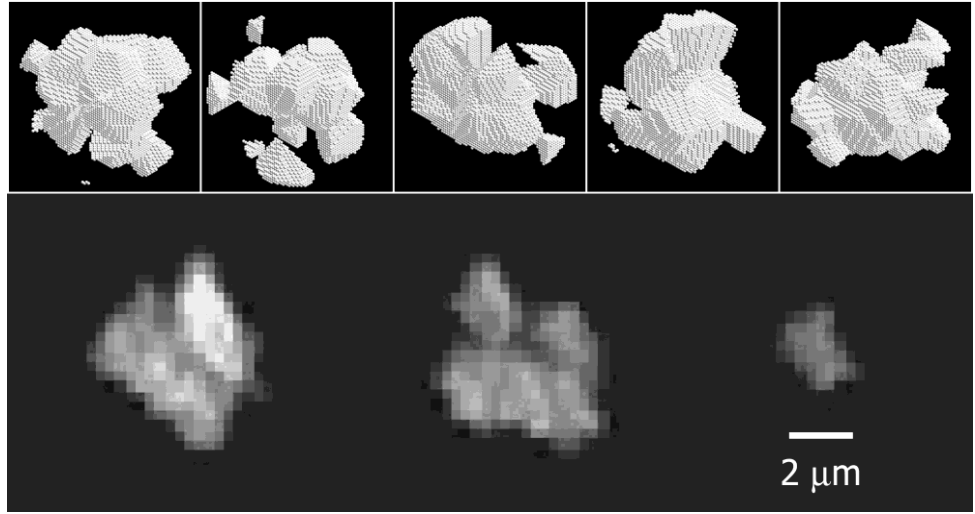

**Figure 4:** Five examples of the agglomerated debris particles (top) and three micron-sized dust particles out of four sampled within the MIDAS experiment aboard *Rosetta* in vicinity of the 67P/Churyumov–Gerasimenko (images adapted from Bentley et al. 2016).

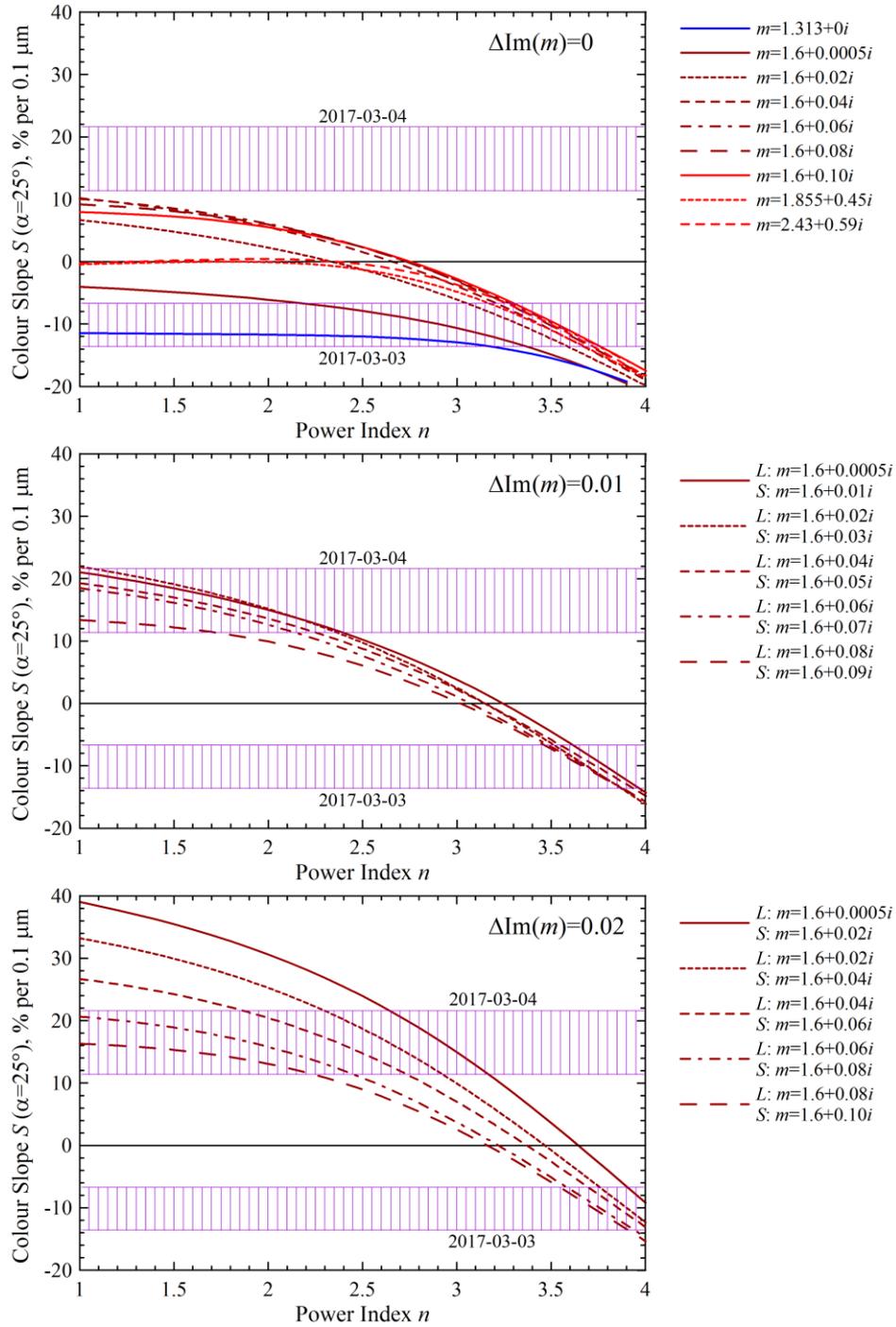

**Figure 5:** Colour slope *S* of agglomerated debris particles as a function of the complex refractive index *m* and index *n* in a power-law size distribution $r^{-n}$. The upper panel shows results for no wavelength-dependence of the imaginary part of refractive index ΔIm(*m*)=0. The middle panel corresponds to the case when the imaginary part at shorter wavelength is larger than at longer wavelength with ΔIm(*m*) = 0.01, and the bottom panel is similar to the middle one, but with ΔIm(*m*) = 0.02.

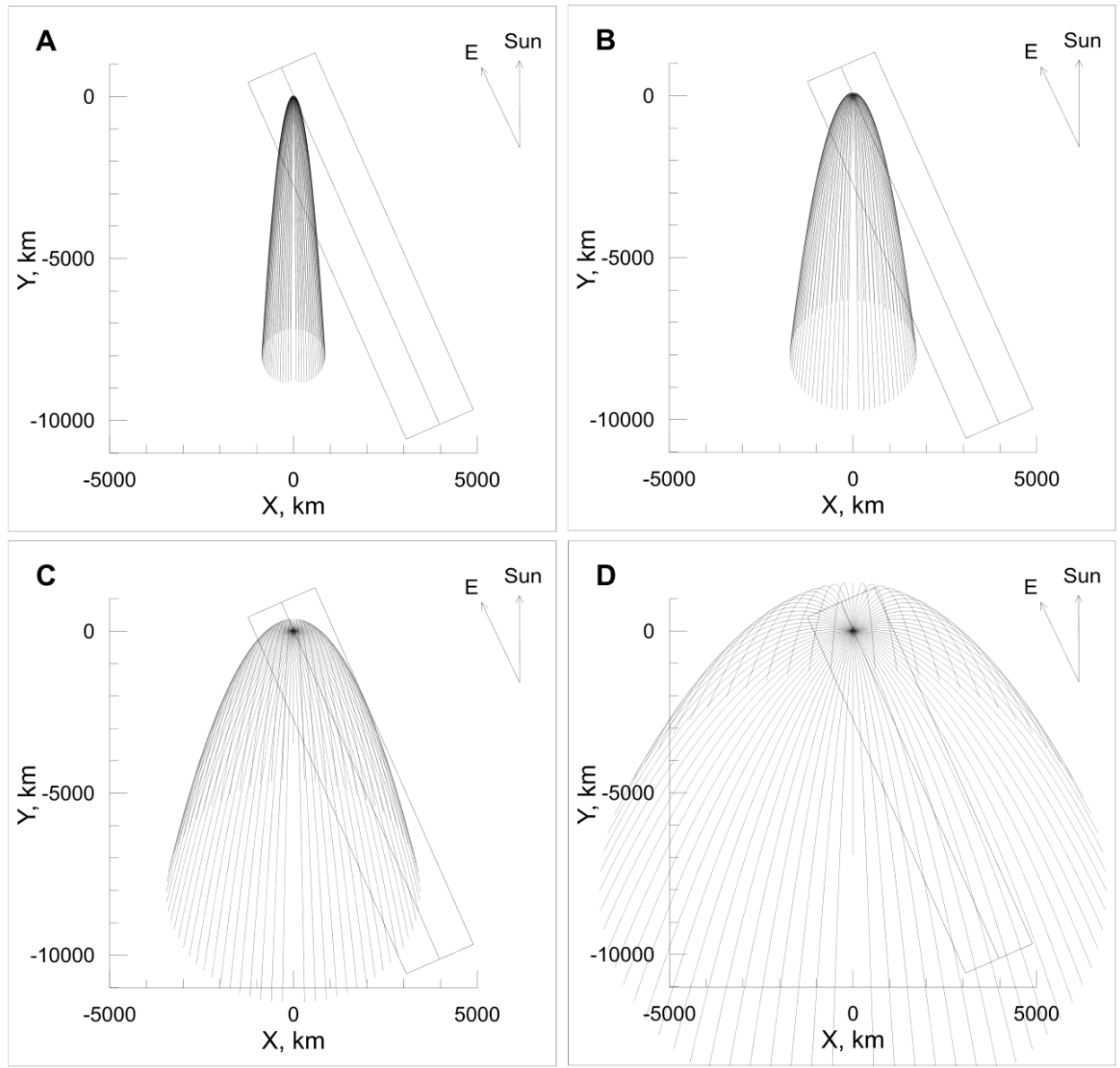

**Figure 6:** Trajectories of the Mg-rich silicate particles with radius $r = 1$ μm and $β = 0.5$ ejected from the 41P/T-G-K nucleus with velocity of 10 m/s (panel A), 20 m/s (B), 40 m/s (C), and 80 m/s (D). The nucleus is placed at the origin of coordinates. Direction to the Sun is upward; whereas, direction to Earth is inclined left for some 24.8° that is shown with an arrow marked E. The short side of the gray rectangular oriented perpendicular to line of sight has length of ~2000 km that corresponds to the aperture in our measurements. Length of trajectories correspond to 1-d flight time of dust particles, except for the case shown in panel D, where some trajectories are clipped on the bottom side of the panel.

**Table 1:** The log of the three-month monitoring of the comet 41P/Tuttle-Giacobini-Kresák

| Date, 2017* | Exposure, sec | Filters & Number** | Seeing, " | Airmass | Observatory Code |
|---|---|---|---|---|---|
| January 29.85 | 240 | V10, R10 | 6.6 | 1.71 | 56 |
| January 30.91 | 60 | V10, R10 | 4.1 | 1.3 | 585 |
| February 15.84 | 100 | V10, R10 | 4.3 | 1.37 | 56 |
| February 15.99 | 60 | V4, R11 | 3.7 | 1.23 | 585 |
| February 25.77 | 100 | V10, R10 | 5.9 | 1.45 | 56 |
| February 26.07 | 30 | V11, R17 | 3.7 | 1.53 | 585 |
| February 27.93 | 30 | V20, R20 | 4.1 | 1.50 | 585 |
| March 03.06 | 60 | V6, R10 | 4.2 | 1.28 | 585 |
| March 04.00 | 60 | V6, R10 | 5.4 | 1.27 | 585 |
| March 14.74 | 40 | V9, R6 | 5.6 | 1.25 | 56 |
| March 20.87 | 100 | V11, R10 | 5.1 | 1.01 | 56 |
| March 24.94 | 14 | V19, R20 | 5.3 | 1.02 | 56 |
| March 25.01 | 30 | V26, R25 | 4.7 | 1.08 | 585 |
| March 26.79 | 14 | V13, R7 | 4.8 | 1.17 | 56 |
| March 27.81 | 14 | V20, R20 | 3.6 | 1.19 | 56 |
| March 27.92 | 30 | V6, R11 | 4.2 | 1.03 | 585 |
| March 28.81 | 14 | V20, R20 | 4 | 1.21 | 56 |
| March 30.82 | 14 | V17, R17 | 4.3 | 1.25 | 56 |
| March 30.93 | 20 | V6, R6 | 3.8 | 1.03 | 585 |
| March 31.83 | 14 | V21, R20 | 4.3 | 1.16 | 56 |
| April 03.08 | 10 | V15, R21 | 4.4 | 1.04 | 585 |
| April 25.01 | 10 | V20, R20 | 3.1 | 1.05 | 585 |

\* Date of start of observations  
\*\* V – V-filter; R – R-filter; number is the number of frames taken

**Table 2:** The results of the three-month monitoring of the colour slope $S$ and the colour variations of comet 41P/Tuttle-Giacobini-Kresák

| Date, 2017* | $r_h$, AU | $\Delta$, AU | $V-R$ | $\alpha,°$ | $S$, % per 0.1 μm |
|---|---|---|---|---|---|
| 29.01 | 1.41 | 0.44 | 0.34 ± 0.07 | 8.7 | -2.42 ± 6.02 |
| 30.01 | 1.41 | 0.43 | 0.33 ± 0.09 | 7.9 | -3.27 ± 7.74 |
| 15.02 | 1.28 | 0.29 | 0.33 ± 0.04 | 5.9 | -3.27 ± 3.44 |
| 25.02 | 1.21 | 0.23 | 0.40 ± 0.01 | 16.8 | 2.75 ± 0.86 |
| 27.02 | 1.19 | 0.22 | 0.38 ± 0.04 | 19.5 | 1.03 ± 3.44 |
| 03.03 | 1.17 | 0.21 | 0.25 ± 0.04 | 24.8 | -10.15 ± 3.43 |
| 04.03 | 1.17 | 0.20 | 0.56 ± 0.05 | 26.0 | 16.48 ± 4.27 |
| 14.03 | 1.12 | 0.17 | 0.36 ± 0.12 | 39.6 | -0.69 ± 10.32 |
| 20.03 | 1.09 | 0.15 | 0.47 ± 0.08 | 48.4 | 8.77 ± 6.87 |
| 24.03 | 1.08 | 0.15 | 0.60 ± 0.07 | 53.9 | 19.89 ± 5.95 |
| 26.03 | 1.07 | 0.14 | 0.51 ± 0.05 | 56.4 | 12.21 ± 4.28 |
| 27.03 | 1.07 | 0.14 | 0.37 ± 0.02 | 57.6 | 0.17 ± 1.72 |
| 28.03 | 1.06 | 0.14 | 0.35 ± 0.04 | 58.8 | -1.55 ± 3.44 |
| 30.03 | 1.06 | 0.14 | 0.42 ± 0.02 | 61.1 | 4.47 ± 1.72 |
| 31.03 | 1.05 | 0.14 | 0.45 ± 0.04 | 62.1 | 7.05 ± 3.44 |
| 03.04 | 1.05 | 0.14 | 0.48 ± 0.02 | 64.3 | 9.63 ± 1.72 |
| 25.04 | 1.06 | 0.18 | 0.41 ± 0.02 | 68.1 | 3.61 ± 1.72 |

* Data of start of observations. Data overnight observations were averaged.